\documentclass[twocolumn,A4]{article}
\usepackage[dvips]{graphics}
\usepackage{setspace}
\usepackage{color}
\definecolor{gold}{rgb}{0.85,0.66,0}
\definecolor{dblue}{rgb}{0,0,0.5}
\begin{document}
\onecolumn
\begin{center}
{\bf{\Large {\textcolor{gold}{Electron transport in polycyclic 
hydrocarbon molecules: A study of shot noise contribution to the 
power spectrum}}}}\\
~\\
{\textcolor{dblue}{Santanu K. Maiti}}$^{1,2,*}$ \\
~\\
{\em $^1$Theoretical Condensed Matter Physics Division,
Saha Institute of Nuclear Physics, \\
1/AF, Bidhannagar, Kolkata-700 064, India \\
$^2$Department of Physics, Narasinha Dutt College,
129, Belilious Road, Howrah-711 101, India} \\
~\\
{\bf Abstract}
\end{center}
We study electron transport in polycyclic hydrocarbon molecules 
attached to two semi-infinite one-dimensional metallic electrodes by 
the use of Green's function formalism. Parametric calculations based 
on the tight-binding framework are given to investigate the transport 
properties through such molecular bridges. In this context we also 
discuss noise power of current fluctuations and focus our attention 
on the shot noise contribution to the power spectrum. The electron 
transport properties are significantly influenced by (a) length of the 
molecule, (b) molecule-electrode interface geometry and (c) molecular 
coupling strength to the electrodes.
\vskip 1cm
\begin{flushleft}
\end{flushleft}
\vskip 4.5in
\noindent
{\bf ~$^*$Corresponding Author}: Santanu K. Maiti

Electronic mail: santanu.maiti@saha.ac.in
\newpage
\twocolumn

\section{Introduction}

The advancements in nanoscience and technologies prompting a growing number
of researchers across multiple disciplines to attempt to devise innovative
ways for decreasing the size and increasing the performance of 
microelectronic circuits. One possible route is based on the idea of using 
molecules and molecular structures as functional devices. During $1970$'s 
Avriam {\em et al.}~\cite{aviram} first studied theoretically the electron 
transport through molecular bridge systems. Later several numerous 
experiments~\cite{metz,fish,reed1,reed2,tali} have been carried out on 
electron transport through molecules placed between two non-superconducting 
electrodes with few nanometer separation. Though in literature both 
theoretical~\cite{baer4,baer5,baer6,baer7,orella1,orella2} as well as 
experimental~\cite{metz,fish,reed1,reed2,tali} 
works on electron transport in several bridge systems are available, yet 
lot of controversies are still present between the theory and experiment, 
and the complete knowledge of the conduction mechanism in this scale is 
not very well established even today. Molecular transport 
properties are characterized by several important factors. First one, of 
course, is the quantization of energy levels associated with the identity 
of the molecule itself and the quantum interference of electron 
waves~\cite{mag,lau,baer1,baer2,baer3,gold,ern1} associated with the 
geometry that the molecule adopts within the junction. Second are the 
different parameters of the Hamiltonian that describe the molecular system, 
the electronic structure of the molecule and the molecular coupling to 
the side attached electrodes. To design molecular electronic devices 
with specific properties, it is important to study structure-conductance 
relationships and in a very recent work Ernzerhof {\em et al.}~\cite{ern2} 
have presented a general design principle and performed several model 
calculations to demonstrate the concept.
The knowledge of current fluctuations (of thermal or quantum origin) also
gives many key ideas for fabrication of efficient molecular devices.
Blanter {\em et al.}~\cite{butt} have described elaborately how the lowest
possible noise power of the current fluctuations can be determined in a
two-terminal conductor. The steady state current fluctuations, the so-called
shot noise, is a consequence of the quantization of charge and it can be 
used to obtain information on a system which is not directly available 
through conductance measurements. The noise power of current fluctuations 
provides an additional important information about the electron correlation 
by calculating the Fano factor ($F$) which directly informs us whether the 
magnitude of the shot noise achieves the Poisson limit ($F=1$) or the 
sub-Poisson ($F<1$) limit.

There exist different {\em ab initio} methods for the calculation of
conductance~\cite{yal,ven,tagami,xue,tay,der,dam} through a molecular 
bridge. At the same time the tight-binding model has been extensively 
studied in the literature and it has also been extended to DFT transport
calculations~\cite{elst}. The study of static density functional theory 
(DFT)~\cite{kohn} within the local-density approximation (LDA) to 
investigate the electronic transport through nanoscale conductors, like 
atomic-scale point contacts, has met with nice success. But, when this 
similar theory applies to molecular junctions, theoretical conductances 
achieve larger values compared to the experimental predictions and these 
quantitative discrepancies need extensive study in this particular field. 
In a recent work, Sai {\em et al.}~\cite{sai} have predicted a correction 
to the conductance using the time-dependent current-density functional 
theory since the dynamical effects give significant contribution in the 
electron transport, and illustrated some important results with specific 
examples. Similar dynamical effects have also been reported in some other 
recent papers~\cite{bush,ven1}, where authors have abandoned the infinite 
reservoirs, as originally introduced by Landauer, and considered two large 
but finite oppositely charged electrodes connected by a nanojunction.
Our aim of the present paper is to reproduce an analytic approach based on 
the tight-binding model to characterize the electronic transport properties
through some polycyclic hydrocarbon molecules and focus our attention on 
the effects of (a) length of the molecules, (b) molecule-to-electrode 
coupling strength and (c) quantum interferences of electronic waves passing
through different arms of the molecular rings. We utilize a simple 
parametric approach~\cite{muj1,muj2,walc1,walc2,sam,hjo} for these 
calculations. The model calculations are motivated by the fact that the 
{\em ab initio} theories are computationally too expensive, while the model 
calculations by using the tight-binding framework are computationally 
very cheap and also provide a worth insight to the problem. In our present 
study attention is drawn on the qualitative behavior of the physical 
quantities rather than the quantitative ones. Not only that, the {\em ab 
initio} theories do not give any new qualitative behavior for this 
particular study in which we concentrate ourselves.

We organize the paper specifically as follows. Section $2$ describes very 
briefly the methodology for the calculation of the transmission probability 
($T$), current ($I$) and noise power of current fluctuations ($S$) 
through a molecule sandwiched between two metallic electrodes by using 
the Green's function formalism. In Section $3$, we study the behavior of 
the conductance as a function of the injecting electron energy, current 
and noise power of its fluctuations as a function of the applied bias 
voltage for the different polycyclic hydrocarbon molecules (Fig.~\ref{poly}). 
Finally, we summarize our results in Section $4$.

\section{Theoretical formulation}

\begin{figure}[ht]
{\centering \resizebox*{7.5cm}{10cm}{\includegraphics{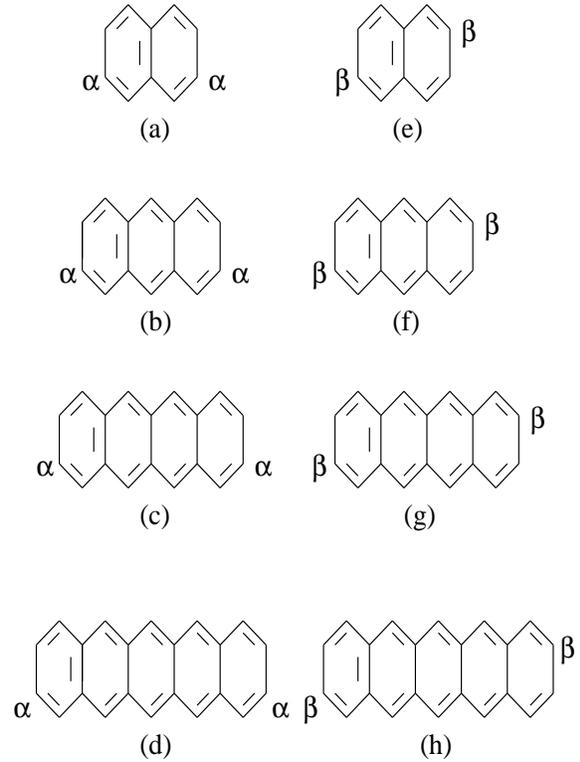}}\par}
\caption{Schematic view of four polycyclic hydrocarbon molecules those are 
attached to two electrodes, namely, source and drain, in {\em cis} 
($\alpha$-$\alpha$) and {\em trans} ($\beta$-$\beta$) configurations by 
thiol (sulfur-hydrogen i.e., S-H bond) groups. These four molecules are 
defined as: napthalene (two rings), anthracene (three rings), tetracene 
(four rings) and pentacene (five rings).}
\label{poly}
\end{figure}

In this section we briefly describe the technique for the calculation of 
transmission probability ($T$), conductance ($g$), current ($I$) and 
noise power of its fluctuations ($S$) through a molecule (schematically 
represented as in Fig.~\ref{poly}) attached to two semi-infinite 
one-dimensional ($1$D) metallic electrodes by using the Green's function 
approach.

At low voltage and low temperature, the conductance $g$ of the molecule is 
given by the Landauer conductance formula~\cite{datta},
\begin{equation}
g=\frac{2e^2}{h} T
\label{equ1}
\end{equation}
where the transmission probability $T$ is written in the form~\cite{datta},
\begin{equation}
T={\mbox{Tr}} \left[\Gamma_S G_M^r \Gamma_D G_M^a\right]
\label{equ2}
\end{equation}
where $G_M^r$ ($G_M^a$) is the retarded (advanced) Green's function of the
molecule and $\Gamma_S$ ($\Gamma_D$) describes its coupling to the source 
(drain). The effective Green's function of the molecule is expressed as,
\begin{equation}
G_M=\left(E-H_M-\Sigma_S-\Sigma_D\right)^{-1}
\label{equ3}
\end{equation}
where $E$ is the energy of the source electron and $H_M$ is the 
Hamiltonian of the molecule which can be written in the tight-binding model 
within the non-interacting picture like,
\begin{equation}
H_M=\sum_i \epsilon_i c_i^{\dagger} c_i + \sum_{<ij>} t \left(c_i^{\dagger}
c_j + c_j^{\dagger} c_i\right)
\label{equ4}
\end{equation}
Here, $\epsilon_i$'s are the site energies and $t$ is the nearest-neighbor
hopping integral. In Eq.(\ref{equ3}), $\Sigma_S$ and $\Sigma_D$ correspond to
the self-energies due to coupling of the molecule to the two electrodes.
Now all the information about the molecule-to-electrode coupling are 
included into these two self-energies and are described through the use 
of Newns-Anderson chemisorption theory~\cite{muj1,muj2}.

The current passing through the molecule can be considered as a single 
electron scattering process between the two reservoirs of charge carriers. 
The current-voltage relation can be obtained from the expression~\cite{datta},
\begin{equation}
I(V)=\frac{e}{\pi \hbar}\int \limits_{-\infty}^{\infty} 
\left(f_S-f_D\right) T(E)~ dE
\label{equ5}
\end{equation}
where the Fermi distribution function $f_{S(D)}=f\left(E-\mu_{S(D)}\right)$
with the electrochemical potentials $\mu_{S(D)}=E_F\pm eV/2$. 
For the sake of simplicity, here we assume that the entire voltage is 
dropped across the conductor-electrode interfaces and this assumption does
not significantly affect the qualitative aspects of the $I$-$V$ 
characteristics. The assumption is based on the fact that the electric 
field inside the molecule, especially for short molecules, seems to have 
a minimal effect on the conductance-voltage characteristics. On the other 
hand, for quite larger molecules and high bias voltage, the electric
field inside the molecule may play a more significant role depending on the
internal structure of the molecule~\cite{tian}, though the effect is too
small.

The noise power of the current fluctuations is calculated from the
relation~\cite{butt},
\begin{eqnarray}
S & = & \frac{2e^2}{\pi \hbar}\int \limits_{-\infty}^{\infty}\left[T(E)
\left\{f_S \left(1-f_S\right) + f_D\left(1-f_D\right) \right\} \right. 
\nonumber \\
 & & + T(E) \left. \left\{1-T(E)\right\}\left(f_S-f_D\right)^2 \right] dE
\label{equ6}
\end{eqnarray}
where the first two terms of this equation correspond to the equilibrium 
noise contribution and the last term gives the non-equilibrium or shot noise
contribution to the power spectrum. By calculating the noise power of the
current fluctuations we can evaluate the Fano factor $F$, which is essential 
to predict whether the shot noise lies in the Poisson or the sub-Poisson 
regime, through the relation~\cite{butt},
\begin{equation}
F=\frac{S}{2 e I}
\label{equ7}
\end{equation}
For $F=1$, the shot noise achieves the Poisson limit where no electron
correlation exists between the charge carriers. On the other hand, for $F<1$,
the shot noise reaches the sub-Poisson limit and it provides the information
about the electron correlation among the charge carriers.

Here, we study the transport properties at much low temperature ($5$ K),
but the qualitative behavior of all the results are invariant up to some
finite temperature ($\sim 300$ K). For simplicity we take the unit $c=e=h=1$
in our present calculations.

\section{Results and their interpretation}

In this section we investigate the characteristic properties of conductance 
as a function of injecting electron energy, current and noise power of its 
fluctuations as a function of the applied bias voltage for four different 
polycyclic hydrocarbon molecules (Fig.~\ref{poly}). 
These molecules are defined as napthalene (two rings), anthracene (three 
rings), tetracene (four rings) and pentacene (five rings), respectively. 
To emphasize the interference effects on the electron transport, we attach 
the electrodes (source and drain) to the molecules in the two distinct 
configurations. One is the so-called {\em cis} configuration, denoted by the
position $\alpha$-$\alpha$ (first column of Fig.~\ref{poly}) and the 
other one is the so-called {\em trans} configuration which is defined by 
the notation $\beta$-$\beta$ (second column of Fig.~\ref{poly}). Contacting 
the molecules to the electrodes in these two different ways we can control 
very nicely the interference conditions of the electronic waves traversing 
through different arms of the molecular rings, and, we will see that the 
electron transport properties are significantly affected by these 
\begin{figure}[ht]
{\centering \resizebox*{8cm}{8.75cm}{\includegraphics{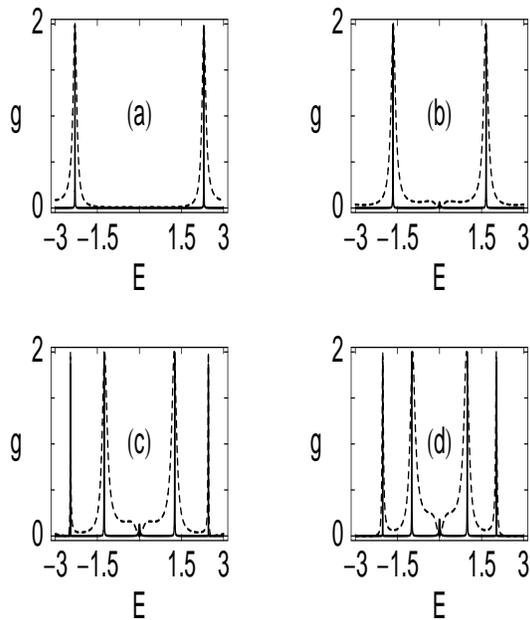}}\par}
\caption{Conductance $g$ as a function of the injecting electron energy 
$E$ for the molecules attached to the electrodes in the {\em cis} 
configuration. (a), (b), (c) and (d) correspond to the results for the
napthalene, anthracene, tetracene and pentacene molecules, respectively. 
The solid and dotted curves denote the weak- and higher-coupling limits, 
respectively.}
\label{ciscond}
\end{figure}
interference effects. In the actual experimental set-up, the electrodes are
generally constructed from gold and the molecules are attached to the 
electrodes via thiol (sulfur-hydrogen i.e., S-H bond) groups in the 
chemisorption technique where the sulfur atoms remove and the hydrogen 
atoms reside.

We will study all the essential features of electron transport through 
the polycyclic hydrocarbon molecules in the two distinct regimes. One is 
the so-called weak-coupling regime defined as $\tau_{S(D)} << t$ and the 
other one is the higher-coupling regime specified as $\tau_{S(D)} 
\sim t$, where $\tau_S$ and $\tau_D$ are the hopping strengths of a
molecule to the source and drain, respectively. In our calculations, 
the parameters for these two distinct regimes are chosen as: 
$\tau_S=\tau_D=0.5$, $t=3$ (weak-coupling) and $\tau_S=\tau_D=2.5$, $t=3$ 
(higher-coupling). Here we take the site energy $\epsilon_i=0$ for all 
the sites of the molecules (for simplicity), and, fix the Fermi energy 
$E_F$ at $0$.

In Fig.~\ref{ciscond}, we plot the conductance ($g$) as a function of the 
injecting electron energy (E) for the molecules attached to the electrodes 
in the {\em cis} configuration, where (a), (b), (c) and (d) correspond to 
the results for the
\begin{figure}[ht]
{\centering \resizebox*{8cm}{8.75cm}{\includegraphics{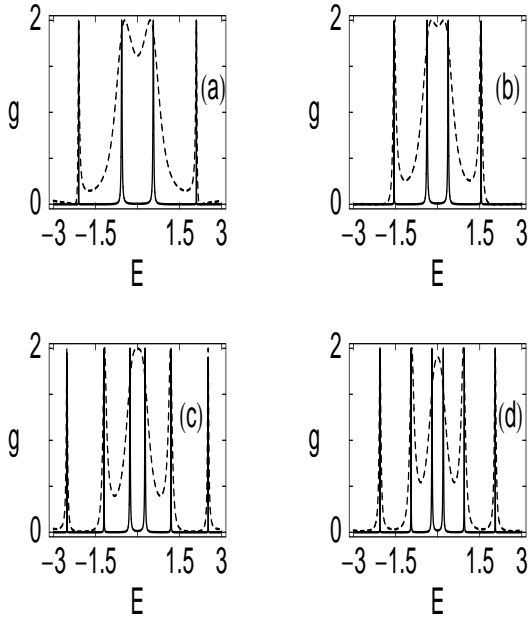}}\par}
\caption{Conductance $g$ as a function of the injecting electron energy 
$E$ for the molecules attached to the electrodes in the {\em trans} 
configuration. (a), (b), (c) and (d) correspond to the results for the
napthalene, anthracene, tetracene and pentacene molecules, respectively.
The solid and dotted curves denote the weak- and higher-coupling limits, 
respectively.}
\label{transcond}
\end{figure}
napthalene, anthracene, tetracene and pentacene molecules, respectively.
The solid curves correspond to the results for the weak-coupling limit,
while the dotted curves denote the results in the limit of higher molecular
coupling. In the weak molecular coupling limit, the conductance shows very
sharp resonant peaks (solid curves in Fig.~\ref{ciscond}) for some specific
energies, while for all other energies it ($g$) almost drops to zero. At 
these resonances the conductance $g$ achieves the value $2$, and accordingly,
the transmission probability $T$ goes to unity, since from the Landauer 
conductance formula we get $g=2T$ (see Eq.(\ref{equ1}) with $e=h=1$ in our 
present calculations). These resonant peaks are associated with the energy
eigenvalues of the corresponding molecule and hence more resonant peaks 
appear with the increment of the length of the molecule. Thus it can be 
emphasized that the conductance spectrum manifests itself the energy 
eigenvalues of the molecule. Now with the increase of the molecular coupling 
strength, the widths of these resonant peaks get enhanced substantially, 
as shown by the dotted curves in Fig.~\ref{ciscond}. This is due to the 
substantial broadening of the molecular energy levels in the limit of
higher molecular coupling. The contribution for this broadening of the 
energy levels comes from the imaginary parts of the self energies 
$\Sigma_S$ and $\Sigma_D$, respectively~\cite{datta}. Therefore, for the 
higher-coupling limit, the electron conducts across the molecules for 
the wide range of energies, while a fine tuning in the energy scale is 
necessary to get the electron conduction through the molecules in the 
limit of weak molecular coupling. Hence, it can be predicted that 
the molecule-to-electrode coupling strength has a significant role in the
determination of the electron conduction through the molecular bridges.

Figure~\ref{transcond} corresponds to the conductance-energy characteristics
for the molecules attached to the electrodes in the {\em trans}
configuration, where (a), (b), (c) and (d) represent the results for the 
napthalene, anthracene, tetracene and pentacene molecules, respectively. 
The solid and dotted curves correspond to the identical meaning as in 
Fig.~\ref{ciscond}. Similar to the case of Fig.~\ref{ciscond}, here we 
also get the sharp resonant peaks in the conductance spectra for the 
weak-coupling limit (solid curves of Fig.~\ref{transcond}), and, they 
(resonant peaks) get broadened in the limit of higher molecular coupling 
(dotted curves of Fig.~\ref{transcond}). The explanations for such 
behaviors are the same as we discuss earlier. From this figure 
(Fig.~\ref{transcond}), it is observed that both for the weak- and 
higher-coupling limits, more resonant peaks appear in the conductance 
spectra i.e., we get more resonating energy eigenstates compared to the 
results obtained for the {\em cis} configuration (Fig.~\ref{ciscond}). 
This feature can be explained as follows. When the molecule is attached to
the electrodes, its energy eigenvalues get modified by the two self-energies
$\Sigma_S$ and $\Sigma_D$, respectively. The self-energies contain the real 
and imaginary parts, where the real part corresponds to the energy shift
and the imaginary part gives the broadening of the molecular energy 
levels~\cite{datta}. By changing the positions of the two electrodes from one
configuration to the other i.e., either from the {\em cis} to the {\em trans} 
configuration or inverse of that, the molecular energy levels are shifted 
in different ways, and accordingly, we get more/less resonating energy
eigenstates in the scale of the energy $E$. This provides different resonant
peaks in the conductance spectra for the two different cases (see 
Figs.~\ref{ciscond} and \ref{transcond}). Another significant observation
is that, in the higher-coupling case the probability amplitude of getting an
\begin{figure}[ht]
{\centering \resizebox*{8cm}{8.75cm}{\includegraphics{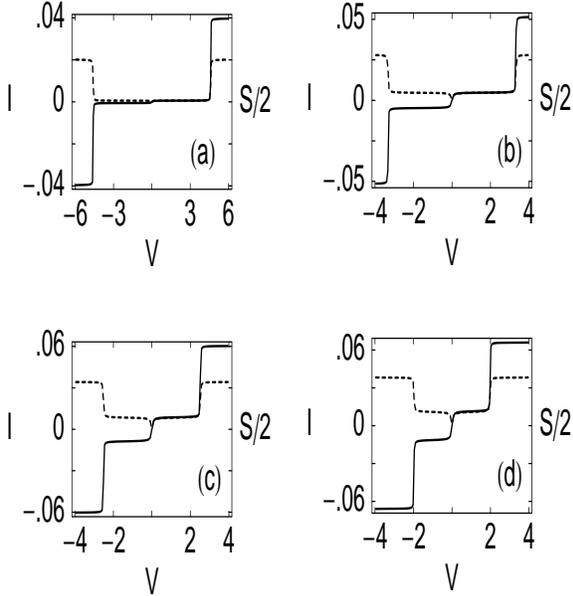}}\par}
\caption{Current $I$ (solid curve) and the noise power of its fluctuations
$S$ (dotted curve) as a function of the applied bias voltage $V$ for the 
molecules attached to the electrodes in the {\em cis} configuration in the 
limit of weak molecular coupling. (a), (b), (c) and (d) correspond to the 
results for the napthalene, anthracene, tetracene and pentacene molecules, 
respectively.}
\label{ciscurrlow}
\end{figure}
electron across the molecular ring in the {\em trans} configuration is much 
greater than that of the {\em cis} configuration for most of the energy values,
except at the resonant peaks where the amplitudes are same ($T=1$) for both 
these two configurations. This is solely due to the quantum interference 
effect of the electronic waves traversing through different arms of the 
molecular rings. This can be explained very simply from the standard 
interpretation of the quantum mechanical theory. The electrons are
carried from the source to drain through the molecules and the electronic
waves propagating along the two arms of the molecular ring may suffer a
relative phase shift between themselves. Accordingly, there might be
constructive or destructive interference due to superposition of the
electronic wave functions along the various pathways. Therefore, the
probability amplitude of the electron across the molecule becomes either
large or small. So electron transmission is strongly affected by the quantum 
interference and can be controlled by the molecule-electrode interface 
geometry. It is observed that the qualitative features of the $g$-$E$ 
\begin{figure}[ht]
{\centering \resizebox*{8cm}{8.75cm}{\includegraphics{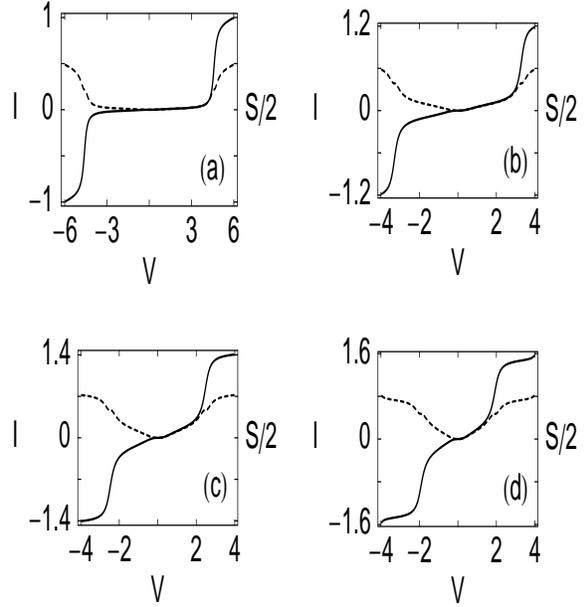}}\par}
\caption{Current $I$ (solid curve) and the noise power of its fluctuations
$S$ (dotted curve) as a function of the applied bias voltage $V$ for the
molecules attached to the electrodes in the {\em cis} configuration in the
limit of higher molecular coupling. (a), (b), (c) and (d) correspond to the 
results for the napthalene, anthracene, tetracene and pentacene molecules, 
respectively.}
\label{ciscurrhigh}
\end{figure}
characteristics are in good agreement with the experimental results. Now 
these interference effects can be visible much more clearly by investigating 
the current-voltage ($I$-$V$) characteristics across the molecules and in 
the forthcoming parts we will describe the current and the noise power of 
its fluctuations as a function of the applied bias voltage for both the 
{\em cis} and the {\em trans} configurations. 

Both the current ($I$) and the noise power of its fluctuations ($S$) are 
calculated from the integration procedure of the transmission function 
($T$), as given in Eqs.(\ref{equ5}) and (\ref{equ6}). The behavior of the
transmission function is exactly similar to that of the conductance 
variation as shown in Figs.~\ref{ciscond} and \ref{transcond}, differ 
only in magnitude by the factor $2$ since we get $g=2T$ from the Landauer 
conductance formula (Eq.(\ref{equ1})). In Fig.~\ref{ciscurrlow}, we plot 
the current and the noise power of its fluctuations as a function of the 
applied bias voltage for the different molecular wires in the limit of 
weak-coupling where the molecules are attached to the electrodes in the 
{\em cis} configuration, where (a), (b), (c) and (d) correspond to the 
results for the napthalene, anthracene, tetracene and pentacene molecules, 
\begin{figure}[ht]
{\centering \resizebox*{8cm}{8.75cm}{\includegraphics{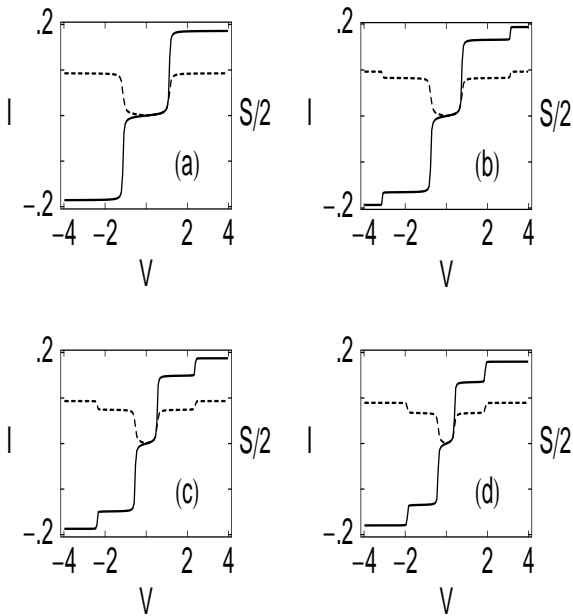}}\par}
\caption{Current $I$ (solid curve) and the noise power of its fluctuations
$S$ (dotted curve) as a function of the applied bias voltage $V$ for the 
molecules attached to the electrodes in the {\em trans} configuration in the 
limit of weak molecular coupling. (a), (b), (c) and (d) correspond to the 
results for the napthalene, anthracene, tetracene and  pentacene molecules, 
respectively.}
\label{transcurrlow}
\end{figure}
respectively. The solid and dotted curves denote the current and noise power, 
respectively. Current shows the staircase-like behavior with sharp steps as 
a function of the applied bias voltage. This is due to the existence of sharp 
resonant peaks in the conductance spectra (see the solid curves
of Fig.~\ref{ciscond}), since the current is evaluated from the integration
method of the transmission function $T$. With the increase of the applied 
bias voltage, the electrochemical potentials on the electrodes are shifted 
gradually and finally cross one of the molecular energy levels. Therefore, 
a current channel is opened up and a jump in the current-voltage 
characteristic curve appears. More steps appear as we increase the length 
of the molecule associated with the molecular resonant states. For all 
these molecular bridges the current amplitudes are very small and they are 
comparable to each other (solid curves of Fig.~\ref{ciscurrlow}). Another 
significant observation is that the threshold bias voltage of the electron 
conduction through the bridge decreases gradually with the increase of the 
length of the molecule. Thus, by controlling the length of the molecule we 
can tune the threshold bias voltage which provides a key idea for 
fabrication of efficient molecular switch. Now in the study of the noise 
power of current fluctuations for these molecular bridges in this 
weak-coupling limit (dotted curves of Fig.~\ref{ciscurrlow}) we see that 
the shot noise goes from the Poisson limit ($F=1$) to the sub-Poisson 
limit ($F<1$) as long as we cross the first step in the current-voltage 
characteristics. This reveals that the electrons are correlated after the 
tunneling process has occurred. Here the electrons are correlated only in the 
sense that one electron feels the existence of the other according to the 
Pauli exclusion principle, since we have neglected all other electron-electron 
interactions in our present formalism. Another important observation is that 
the amplitudes of the noise power in the sub-Poisson region for all these 
bridges are almost invariant. These results emphasize that in the limit of
weak-coupling the noise power of the current fluctuations within the 
sub-Poisson limit remains in the same level independent of the length of 
the molecule.

The characteristic properties of the current and noise power of its 
fluctuations are also very interesting for these molecular bridges 
(molecules attached to the electrodes in the {\em cis} configuration) in 
the limit of higher molecular coupling. The results are represented in 
Fig.~\ref{ciscurrhigh}, where (a), (b), (c) and (d) correspond to the 
results for the same molecular bridges as given in Fig.~\ref{ciscurrlow}, 
respectively. The solid and dotted curves denote the same meaning as in 
Fig.~\ref{ciscurrlow}. It is observed that the current varies quite 
continuously (solid curves of Fig.~\ref{ciscurrhigh}) with the applied bias 
voltage $V$. This is due to the broadening of the resonant peaks in this 
higher-coupling limit (see the dotted curves of Fig.~\ref{ciscond}), as the 
current is computed from the integration procedure of the function $T$. 
The key observation is that the current amplitudes get enhanced quite
significantly compared to the current amplitudes observed for the
molecular bridges in the weak-coupling limit (see the solid curves of
Fig.~\ref{ciscurrlow}). This can be emphasized very simply by noting the
areas under the curves in the conductance spectra for these two coupling 
cases (Fig.~\ref{ciscond}). It is also noted here that the current amplitude
increases gradually as we increase the length of the molecule for this
higher-coupling limit. Now in the determination of the noise power of the
current fluctuations, we see that the shot noise makes a transition from the
\begin{figure}[ht]
{\centering \resizebox*{8cm}{8.75cm}{\includegraphics{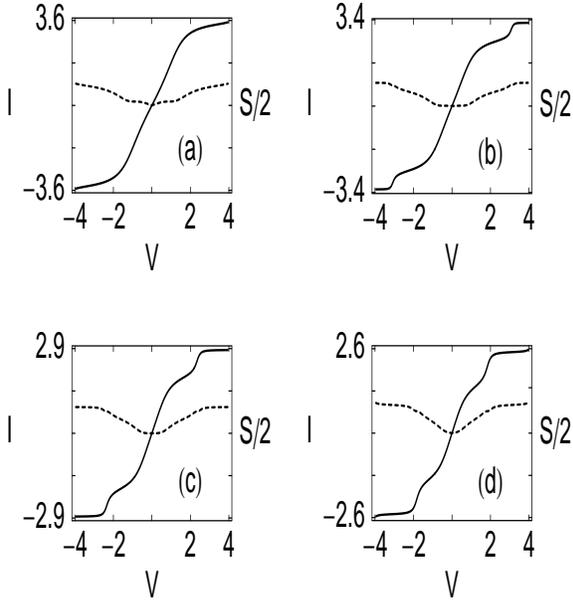}}\par}
\caption{Current $I$ (solid curve) and the noise power of its fluctuations
$S$ (dotted curve) as a function of the applied bias voltage $V$ for the 
molecules attached to the electrodes in the {\em trans} configuration in the 
limit of higher molecular coupling. (a), (b), (c) and (d) correspond to the 
results for the napthalene, anthracene, tetracene and pentacene molecules, 
respectively.}
\label{transcurrhigh}
\end{figure}
Poisson limit ($F=1$) to the sub-Poisson limit ($F<1$) after some critical
value of the applied bias voltage (see the dotted curves of 
Fig.~\ref{ciscurrhigh}). This critical value of the bias voltage decreases
gradually as we increase the length of the molecule so that the shot noise
goes faster to the sub-Poisson limit for the longer molecules. Therefore, 
it predicts that the electron correlation is much more significant for the 
molecules with higher length than that of the shorter ones. In this limiting 
case we also see that, like as in the weak-coupling case, the amplitudes of 
the noise power in the sub-Poisson regime for all these bridges are almost 
comparable to each other.

Now focus our attention on the current and noise power of its fluctuations 
for the molecules attached to the electrodes in the {\em trans} configuration. 
In Fig.~\ref{transcurrlow}, we plot the results for the molecules attached 
to the electrodes in the {\em trans} configuration in the limit of weak 
molecular coupling, where (a), (b), (c) and (d) correspond to the results 
for the napthalene, anthracene, tetracene and pentacene molecules, 
respectively. The solid and dotted curves represent the same meaning as in 
Figs.~\ref{ciscurrlow} and \ref{ciscurrhigh}. Like as in the {\em cis} 
configuration, for the weak-coupling limit here we also see that the 
current shows staircase-like behavior with sharp steps (see the solid 
curves of Fig.~\ref{transcurrlow}) as a function of the applied bias 
voltage and the shot noise makes a transition from the Poisson limit to 
the sub-Poisson limit as we cross the first step in the $I$-$V$ curves 
(see the dotted curves of Fig.~\ref{transcurrlow}). The explanations of 
such behaviors are the same as before. The strange observation is that 
the current amplitudes for all these molecular wires get enhanced quite 
significantly, though we are in the weak-coupling limit, compared to the 
current amplitudes obtained for these molecules in the {\em cis} 
configuration. This is solely due to the quantum interference effects of 
the electronic waves traversing through the different arms of the molecular 
rings (Fig.\ref{poly}). Thus interference effects play significant role in 
the determination of the current through the molecular bridges. Here also, 
like as in Fig.~\ref{ciscurrlow}, the noise power of the current 
fluctuations lies in the same level within the sub-Poisson limit (see the
dotted curves of Fig.~\ref{transcurrlow}), independent of the length of the
molecule.

Figure~\ref{transcurrhigh} displays the results for the molecular bridges
(molecules attached to the electrodes in the {\em trans} configuration)
in the limit of higher-coupling, where (a), (b), (c) and (d) correspond to 
the same molecules as in Fig.~\ref{transcurrlow}. The current varies 
continuously (solid curves of Fig.~\ref{transcurrhigh}) with the bias 
voltage $V$ for all these bridges, similar to the results studied earlier 
for the bridges in the {\em cis} configuration (solid curves of 
Fig.~\ref{ciscurrhigh}). For this higher-coupling limit the current 
amplitudes are also very large than the results obtained for the bridges 
in the {\em cis} configuration with higher molecular coupling and the 
explanation for such kind of behavior is the same as we describe earlier. 
Here it is observed that the current amplitude decreases gradually as we 
increase the length of the molecule, in contrary to the results predicted 
in Fig.~\ref{ciscurrhigh}. Our calculations of the noise power of the 
current fluctuations for such bridges provide that there is no such 
possibility of transition from the Poisson limit to the sub-Poisson limit 
since the shot noise already achieves the sub-Poisson limit (see the dotted 
curves of Fig.~\ref{transcurrhigh}), momentarily as we switch on the bias 
voltage. Accordingly, for such bridges in this limit of molecular coupling 
the electron correlation is highly significant. Thus we can predict that 
the noise power of the current fluctuations strongly depends on the 
molecule-to-electrode interface geometry as well as the 
molecule-to-electrode coupling strength.

\section{Concluding remarks}

In summary of this article, we have introduced parametric calculations based 
on the tight-binding model to investigate the electron transport properties
through some polycyclic hydrocarbon molecules attached to two metallic 
electrodes. We have shown that the electron transport properties are 
significantly affected by (a) length of the molecules, (b) molecule-electrode 
interface geometry and (c) molecule-to-electrode coupling strength. All 
the results described here provide several key ideas for fabrication of 
efficient molecular devices.

The conductance shows fine resonant peaks for the weak-coupling limit 
(solid curves of Figs.~\ref{ciscond} and \ref{transcond}), while they
get broadened in the limit of higher molecular coupling (dotted curves of
Figs.~\ref{ciscond} and \ref{transcond}). This increment of the resonant
widths is due to the broadening of the molecular energy levels, where the
contribution comes from the imaginary parts of the self energies $\Sigma_S$
and $\Sigma_D$~\cite{datta}.

The quantum interference effects on the electron transport have been 
clearly described by studying the current-voltage characteristics. Both 
for the {\em cis} and {\em trans} configurations, in the limit of weak 
molecular coupling, the current gets the staircase-like behavior with 
sharp steps (solid curves of Figs.~\ref{ciscurrlow} and \ref{transcurrlow}). 
On the other hand, it varies quite continuously with the applied bias 
voltage in the limit of higher molecular coupling (solid curves of 
Figs.~\ref{ciscurrhigh} and \ref{transcurrhigh}). The most remarkable 
observation is that by increasing the molecular coupling strength one can 
enhance the current amplitude quite significantly. We can also tune the 
current amplitude in a controllable way by varying the molecule-to-electrode 
interface geometry associated with the quantum interference effects. 

In the study of the noise power of current fluctuations, we have seen that
it strongly depends on the molecule-to-electrode interface geometry as well 
as the molecular coupling strength. In the weak-coupling case both 
for the {\em cis} and {\em trans} configurations, the shot noise makes 
a transition from the Poisson limit ($F=1$) to the sub-Poisson limit ($F<1$) 
as we cross the first step in the $I$-$V$ characteristics (dotted curves of 
Figs.~\ref{ciscurrlow} and \ref{transcurrlow}) which reveals that the 
electrons are correlated with each other after the tunneling process has 
occurred. Quite similarly in the higher-coupling limit for the molecular 
bridges in the {\em cis} configuration, the shot noise makes a transition 
from the Poisson limit to the sub-Poisson limit beyond some critical value 
of the applied bias voltage $V$ (dotted curves of Fig.~\ref{ciscurrhigh}). 
On the other hand, for the molecules attached to the electrodes in the
{\em trans} configuration and in the limit of higher-coupling, there is no
such possibility of transition from the Poisson limit to the sub-Poisson 
limit as the shot noise already lies in the sub-Poisson limit momentarily 
as we switch on the bias voltage (dotted curves of Fig.~\ref{transcurrhigh}), 
and accordingly, for such cases the electron correlation is very much 
significant. 

Throughout our study we have used several important approximations by 
neglecting the effects of electron-electron interaction, all the 
inelastic scattering processes, Schottky effect, static Stark effect, 
etc. More studies are expected to take into account all these 
approximations for our further investigations.

\end{document}